\documentclass[aps,prb,twocolumn,superscriptaddress,showpacs]{revtex4}
\usepackage{dcolumn} 
\usepackage{graphicx}
\usepackage{dcolumn} 
\usepackage{bm}      
\usepackage{harpoon}
\usepackage{color}
\usepackage{amssymb}
\usepackage{amsfonts}
\usepackage{mathrsfs}
\usepackage{mathtools}
\usepackage{array}
\usepackage{amsmath,amssymb}
\usepackage{multirow}
\usepackage{epstopdf}
\usepackage{latexsym}
\usepackage{subfigure}
\usepackage{dsfont} 
\usepackage{wasysym} 
\usepackage{times}
\usepackage{dcolumn} 
\usepackage{bm}      
\usepackage{epstopdf}
\usepackage{gensymb}
\usepackage{textcomp}

\begin{document}

\title{Magnetic Properties of  the low dimensional BaM$_2$Si$_2$O$_7$(M= Cu, Co, Mn) system}

\author{G. H. Wang}
\affiliation{Key laboratory of Artificial Structures and Quantum Control, School of Physics and Astronomy, Shanghai JiaoTong University, Shanghai, 200240, China}

\author{C. Y. Xu}
\affiliation{Key laboratory of Artificial Structures and Quantum Control, School of Physics and Astronomy, Shanghai JiaoTong University, Shanghai, 200240, China}

\author{H. B. Cao}
\affiliation{Quantum Condensed Matter Division, Oak Ridge National Laboratory, Oak Ridge, Tennessee, 37831, USA.}

\author{T. Hong}
\affiliation{Quantum Condensed Matter Division, Oak Ridge National Laboratory, Oak Ridge, Tennessee, 37831, USA.}

\author{Q. Huang}
\affiliation{Department of Physics and Astronomy, University of Tennessee, Knoxville, Tennessee 37996-1200, USA}

\author{Q. Y. Ren}
\affiliation{Key laboratory of Artificial Structures and Quantum Control, School of Physics and Astronomy, Shanghai JiaoTong University, Shanghai, 200240, China}

\author{J. Q. Xu}
\affiliation{Key laboratory of Artificial Structures and Quantum Control, School of Physics and Astronomy, Shanghai JiaoTong University, Shanghai, 200240, China}

\author{H. D. Zhou}
\affiliation{Key laboratory of Artificial Structures and Quantum Control, School of Physics and Astronomy, Shanghai JiaoTong University, Shanghai, 200240, China}
\affiliation{Department of Physics and Astronomy, University of Tennessee, Knoxville, Tennessee 37996-1200, USA}

\author{W. D. Luo}
\affiliation{Key laboratory of Artificial Structures and Quantum Control, School of Physics and Astronomy, Shanghai JiaoTong University, Shanghai, 200240, China}
\affiliation{Collaborative Innovation Center of Advanced Microstructures, Nanjing, 210093, China}
\affiliation{Institute of Natural Sciences, Shanghai Jiao Tong University, Shanghai, 200240, China}

\author{D. Qian}
\affiliation{Key laboratory of Artificial Structures and Quantum Control, School of Physics and Astronomy, Shanghai JiaoTong University, Shanghai, 200240, China}
\affiliation{Collaborative Innovation Center of Advanced Microstructures, Nanjing, 210093, China}

\author{J. Ma}
\email{jma3@sjtu.edu.cn}
\affiliation{Key laboratory of Artificial Structures and Quantum Control, School of Physics and Astronomy, Shanghai JiaoTong University, Shanghai, 200240, China}
\affiliation{Collaborative Innovation Center of Advanced Microstructures, Nanjing, 210093, China}

\date{\today}

\begin{abstract}
We performed susceptibility, magnetization, specific heat, and single crystal neutron diffraction measurements on single crystalline BaMn$_2$Si$_2$O$_7$. Based on the results, we revisited its spin structure with a more accurate solution and constructed a magnetic phase diagram with applied field along the $b$-axis, which contains a spin flop transition around 6 T. We also used susceptibility, magnetization, and specific heat results confirmed the ferrimagnetic-like magnetism in polycrystalline  BaCo$_2$Si$_2$O$_7$. Furthermore, we performed LSDA + U calculations for the BaM$_2$Si$_2$O$_7$ (M = Cu, Co, and Mn) system. Our discussions based on the comparison among the obtained magnetic exchange interactions suggest the different structures and electronic configurations are the reasons for the different magnetic properties among the system members.
\end{abstract}
\maketitle

\section{INTRODUCTION}
The quasi one-dimensional (1D) magnetic system has received considerable attention due to the unique magnetic properties. It has been proved in theory that there is no magnetic order formed even at 0 K in a purely one-dimensional spin 1/2 chain system but there will be a Neel ordering state if a three-dimensional(3D) antiferromagnet is present due to the appearance of the inter-chain interactions\cite{Ising,Heisenberg,Mermin,Kim}. Such a quasi 1D system with the week coupling between the magnetic chains will exhibit a crossover from 1D magnetic behavior \cite{Griffiths, Bonner, Ground energy, 1D anisotropy, spin perils, Eggert}at high temperatures to a 3D ordered state at the low temperatures. Moreover, such a system with strong spin anisotropy usually exhibit rich phase diagram under applied magnetic field below the ordering temperature.

\begin{figure}[tbp]
\setlength{\abovecaptionskip}{-0.1cm}
	\linespread{1}
	\par
	\begin{center}
		\includegraphics[width= 3.2 in]{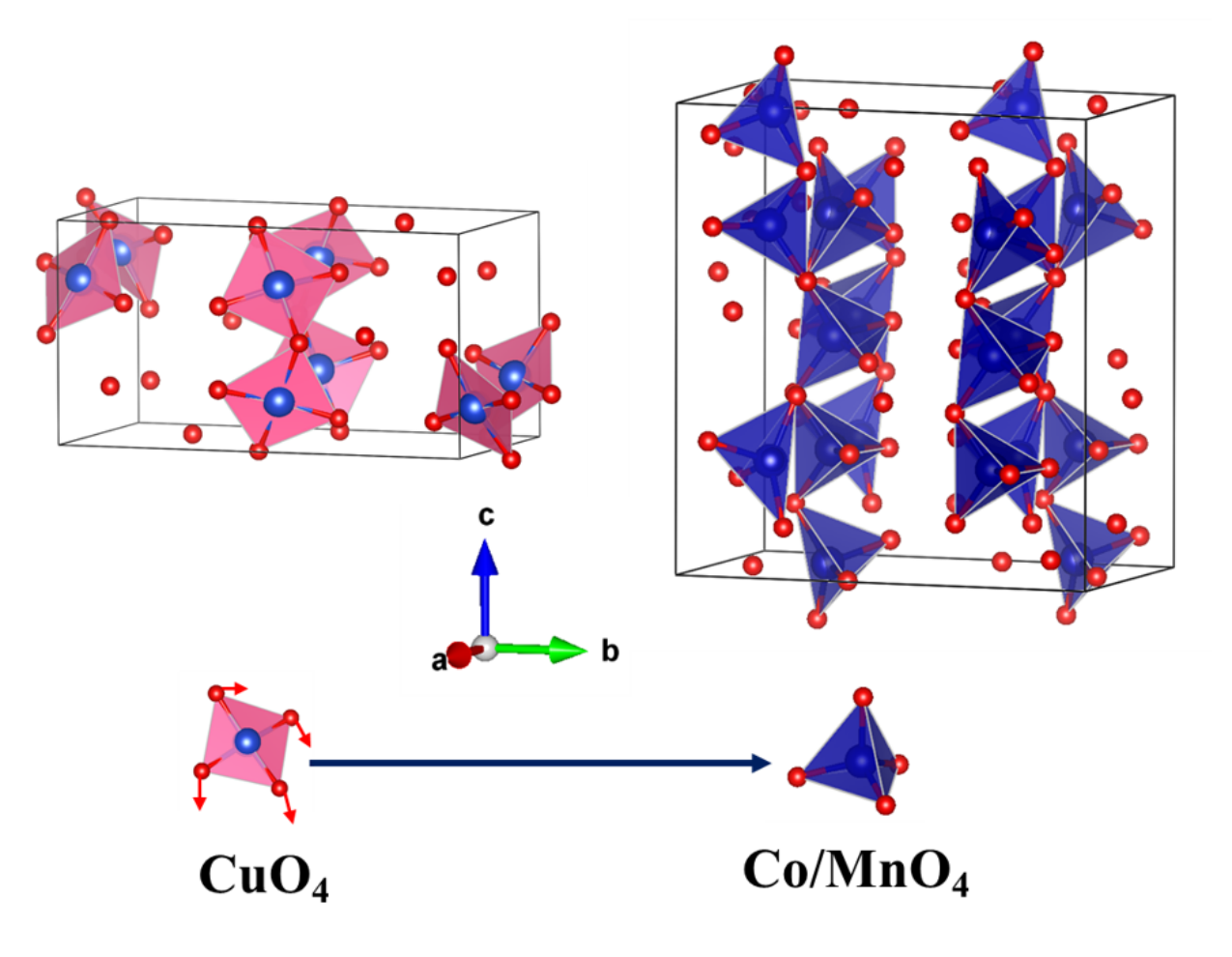}
	\end{center}
	\par
	\caption{(color online) a) Schematic crystalline structures for (a) BaCu$_2$Si$_2$O$_7$ and (b) BaM$_2$Si$_2$O$_7$ (M = Mn and Co)
	}
\end{figure}

The BaM$_2$Si$_2$O$_7$  (M = Cu, Co, and Mn) with layer structure is an excellent system to study such low-dimensional magnetic behaviors. In this system, the BaCu$_2$Si$_2$O$_7$  has been extensively studied due to its spin 1/2 (Cu$^{2+}$) nature. This compound exhibits an orthorhombic structure with the space group of Pnma. The nearest Cu$^{2+}$ ions through the O$^{2-}$ ions construct a zigzag chain along the $c$-axis. Below $T_{\text{N}}$ = 9.2 K, an antiferromagnetic ordered state is formed, in which the magnetic easy axis is the $c$-axis\cite{Tsukada}. With applied magnetic field along the $c$-axis, a multiple step spin flop transitions have been reported\cite{Takeya,Ressouche,Glazkov,Dhalenne}. Moreover, its exchange interactions have been studied by inelastic neutron scattering measurements and theoretical calculations\cite{Zheludev1,Kenzelmann,Zheludev,Hayn,Raymond}.

The replacement of  Cu$^{2+}$ ions with larger size Co$^{2+}$ and Mn$^{2+}$ ions, a structural distortion is achieved. Now, BaCo$_2$Si$_2$O$_7$ and BaMn$_2$Si$_2$O$_7$ have a monoclinic crystal structure with the space group C2/c\cite{Richard}. Accordingly, the M-O ligands for the CuO$_4$ plaquettes in the Cu sample now form a  Co/MnO$_4$ tetrahedron in the Co and Mn samples\cite{Layland,Lu}, as shown in Fig. 1. For BaCo$_2$Si$_2$O$_7$ wtih Spin-3/2 (Co$^{2+}$), very limited study suggested a ferromagnetic-like ground state\cite{Layland,Akaki}. For  BaMn$_2$Si$_2$O$_7$, the powder neutron diffraction studies presented a possible spin structure below $T_{\text{N}}$ = 26 K\cite{J. Ma}. A simple magnetization measured with applied field along the $b$-axis suggested a spin flop transition. Therefore, it is obvious that all three samples have distinct magnetic properties for the BaM$_2$Si$_2$O$_7$ system.

Due to the limited studies on the Co and Mn samples, several questions are left open. For example, the ground state of Co sample needs to be confirmed by more detailed studies. For Mn sample, although the spin flop transition was reported but what is its exact magnetic phase diagram? Its spin structure was solved based on polycrystalline studies and certainly deserves a revisit by single crystal neutron diffraction experiment, which usually leads to more accurate spin structure solutions. Most importantly, so far there is no systematical studies on this system to better understand the driving force for their distinct magnetic ground states. In this paper, first, we  performed detailed magnetic properties, specific heat, and single crystal neutron diffraction measurements under applied fields to revisit the spin structure and construct the magnetic phase diagram for single crystalline BaMn$_2$Si$_2$O$_7$; second, we studied the physical properties of polycrystalline BaCo$_2$Si$_2$O$_7$ to confirm its magnetic ground state; third, we performed LSDA + U calculations to systematically study the exchange interactions of the system. Our discussions based on the detailed comparison among the obtained exchange interactions for the system point out that the structure distortion and the out shell electronic configuration are two main reasons for the systematical changes of the magnetic ground states for these three samples.

\section{EXPERIMENTAL}

The single crystal of BaMn$_2$Si$_2$O$_7$ was grown by using an image furnace. The detailed information is described in Ref.\cite{J. Ma}. The polycrystalline BaCo$_2$Si$_2$O$_7$ was synthesized by solid state reactions. The  stoichiometric mixtures of BaCO$_3$, CoCO$_3$, and SiO$_2$ were ground and reacted at a temperature of 1100 $^{\circ}$C in air for 60 hours with several intermediate grindings. The magnetic susceptibility, magnetization, and heat capacity were measured using a  PPMS (physical property measurement system, Quantum Design). The single-crystal neutron diffraction was performed at the HB-3A Four-Circle Diffractometer and triple axis spectrometers CG-4C at the High Flux Isotope Reactor at Oak Ridge National Laboratory. A neutron wavelength of 1.005 \AA~from a bent perfect Si-113 monochromator was used in HB-3A and the collimation of open-open-80'-open was utilized in CG-4C. The neutron diffraction data was refined by the Rietveld refinement program FullProf\cite{J. Rodr}.

\section{RESULTS}
\subsection{BaMn$_2$Si$_2$O$_7$}

\begin{figure}[tbp]
\setlength{\abovecaptionskip}{-0.1cm}
	\linespread{1}
	\par
	\begin{center}
		\includegraphics[width= 3.2 in]{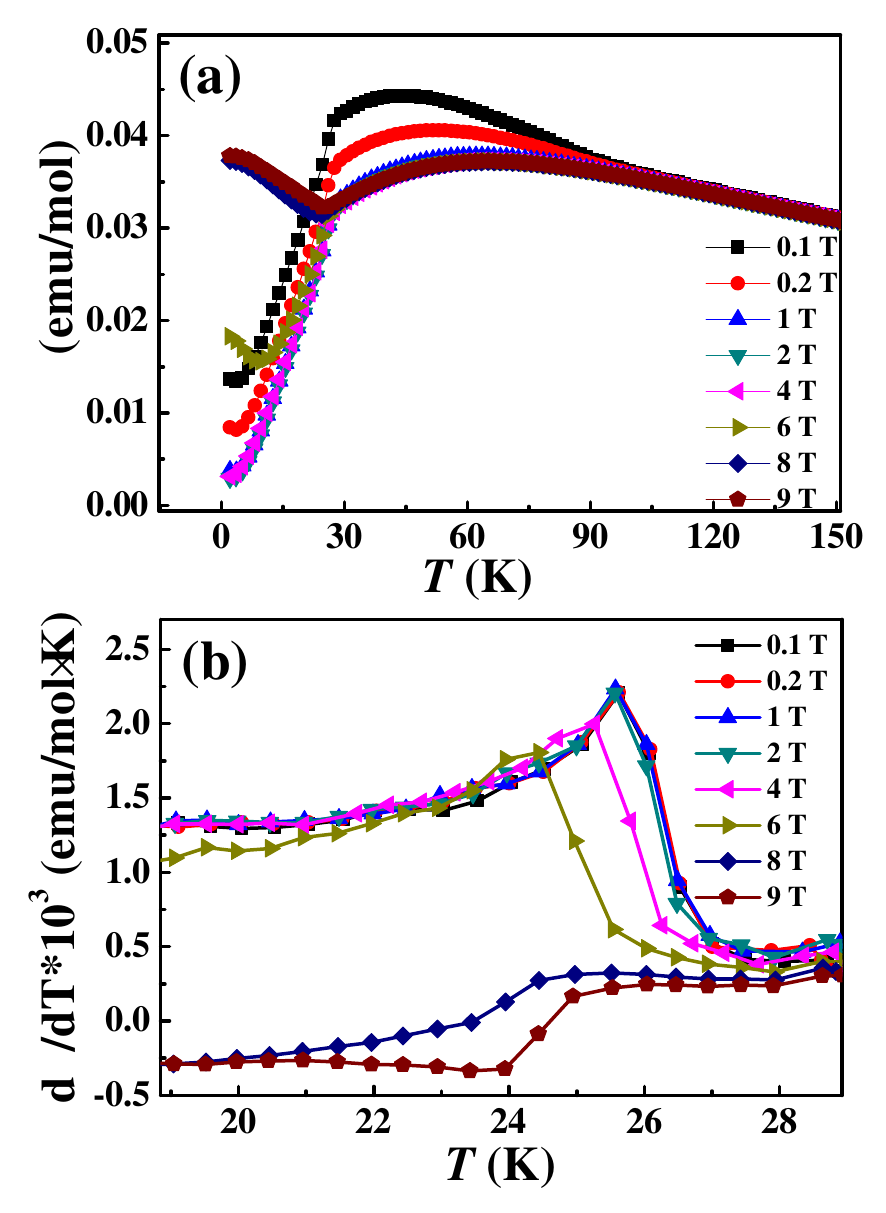}
	\end{center}
	\par
	\caption{(color online) (a) The temperature dependence of susceptibility ($\chi$) (a) and its derivative, d$\chi$/d$T$ (b) at selected fields for BaMn$_2$Si$_2$O$_7$. The applied field is along the $b$-axis.
}
\end{figure}

The temperature dependence of susceptibility, $\chi$,  under different magnetic fields for BaMn$_2$Si$_2$O$_7$ is shwon in Fig. 2. With applied field H = 0.1 T along the $b$-axis, $\chi$ shows a broad peak around 30 $\sim$ 40 K, and then a sharp decline around 26 K with decreasing temperatures. This data suggests that BaMn$_2$Si$_2$O$_7$ undergoes a transition from the two dimensional short range ordering related to the broad peak to a long range magnetic ordering related to the sharp drop. Similar behavior has also been observed for other Heisenberg antiferromagnetic (AFM) linear-chain systems\cite{Dingle,Maeshima}. Here, the ordering temperature, $T_{\text{N}}$ = 26 K for 0.1 T data,  is defined as the peak position of the derivative of $\chi$, as shown in Fig. 2(b). For H $\leq$ 6 T, $T_{\text{N}}$ decreases with increasing fields. Meanwhile for H $\geq$ 8 T, instead of decrease, $\chi$ now increases below $T_{\text{N}}$ with decreasing temperatures. Moreover, $T_{\text{N}}$ now increases with increasing fields. This distinct behavior between different field regimes suggests that there is a kind of spin state transition, or spin flop transition, induced by applied field around 6 $\sim$ 8 T.

\begin{figure}[tbp]
\setlength{\abovecaptionskip}{-0.1cm}
	\linespread{1}
	\par
	\begin{center}
		\includegraphics[width= 3.2 in]{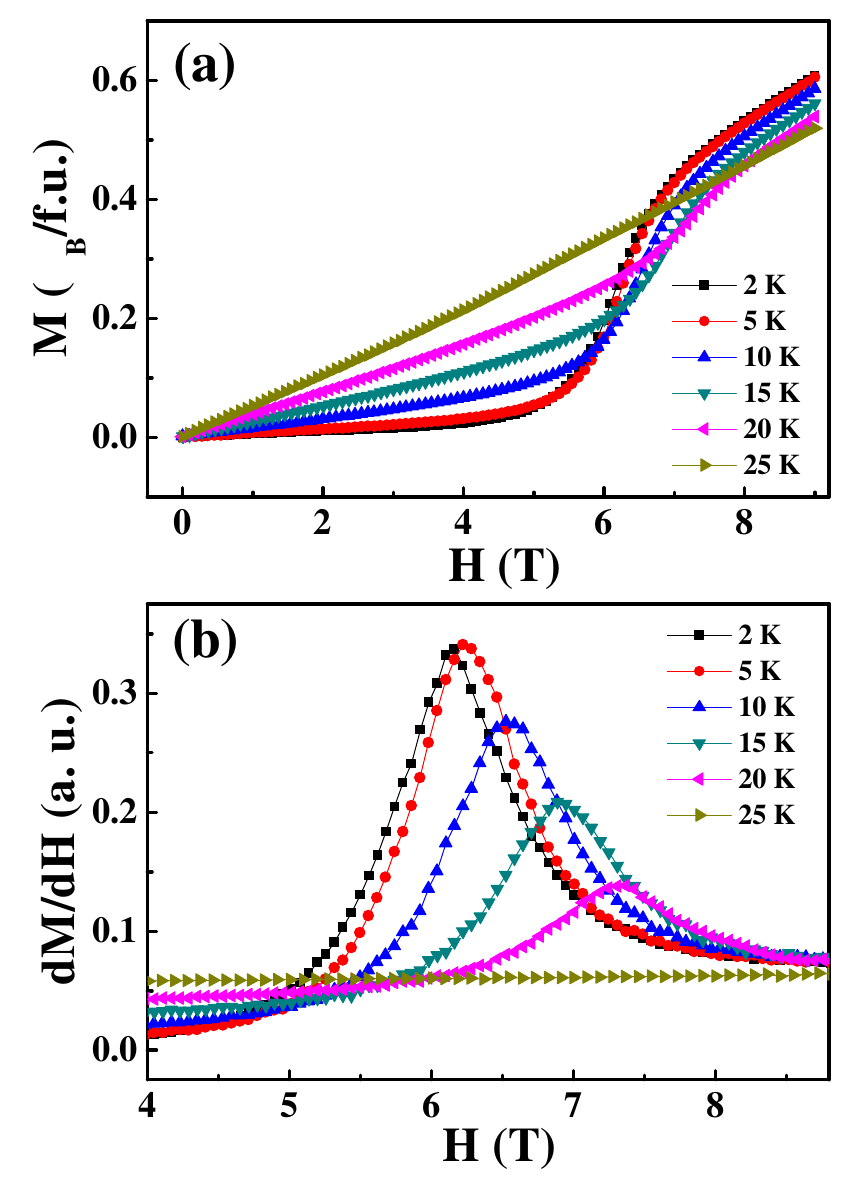}
	\end{center}
	\par
	\caption{(color online) (a) The field dependence of magnetization (M) (a) and its derivative, dM/dH (b) at selected temperatures for BaMn$_2$Si$_2$O$_7$ The applied field is along the $b$-axis.
}
\end{figure}

The field dependence of magnetization, M, at selected temperatures for BaMn$_2$Si$_2$O$_7$ is shown in Fig. 3. With H along the $b$-axis and temperatures below $T_{\text{N}}$, M exhibits a sharp slope change around 6 T. Accordingly, the dM/dH curve (Fig. 3(b)) shows a sharp peak, which position is defined as the critical field H$_{\text{c}}$. At 2 K, H$_{\text{c}}$ = 6.0 T and it shifts to higher fields with increasing temperatures. For temperature near the $T_{\text{N}}$, such as 25 K, this increase becomes very weak. This critical field should be the one for inducing the spin flop transition.

\begin{figure}[tbp]
\setlength{\abovecaptionskip}{-0.1cm}
	\linespread{1}
	\par
	\begin{center}
		\includegraphics[width= 3.2 in]{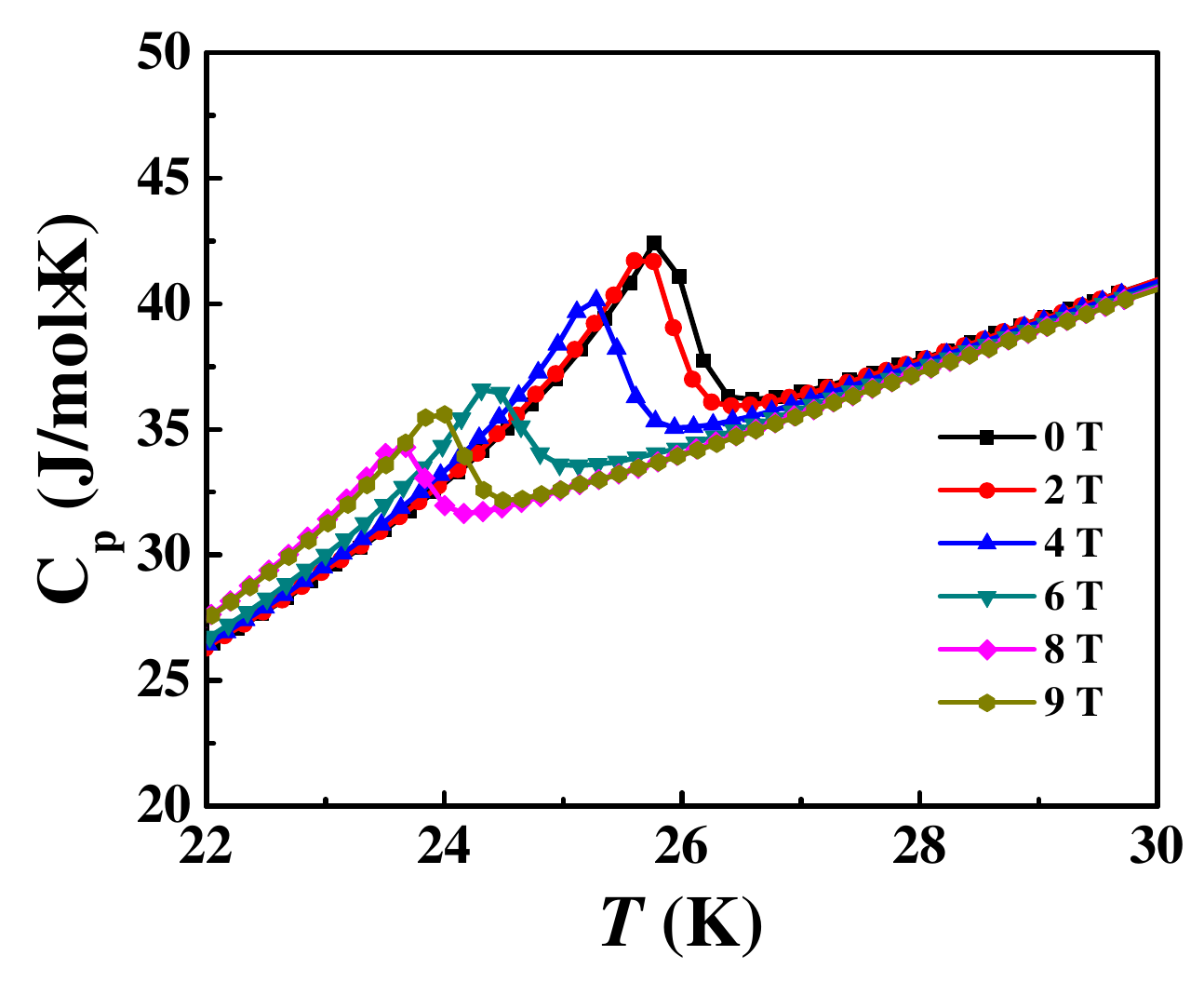}
	\end{center}
	\par
	\caption{(color online) The temperature dependence of specific heat around the transitions at different fields for BaMn$_2$Si$_2$O$_7$ The applied field is along the $b$-axis.
}
\end{figure}

The specific heat for BaMn$_2$Si$_2$O$_7$ was plotted in Fig. 4. At zero filed, the data shows a sharp peak at 26 K, which is corresponding to the $T_{\text{N}}$. For H $\leq$ 6 T along the $b$-axis, the peak (or $T_{\text{N}}$) shifts to lower temperatures with increasing fields. However for H = 9 T data, its peak position is higher than the 8 T data, which is consistent with the trend of $T_{\text{N}}$ defined by the susceptibility data. Therefore, the susceptibility, magnetization, and specific heat data all consistently exhibit a field induced spin flop transition around 6 T for BaMn$_2$Si$_2$O$_7$. Similar spin flop transition also has been observed in BaCu$_2$Si$_2$O$_7$\cite{Takeya}.

 \begin{figure}[tbp]
 \setlength{\abovecaptionskip}{-0.1cm}
	\linespread{1}
	\par
	\begin{center}
		\includegraphics[width= 3.2 in]{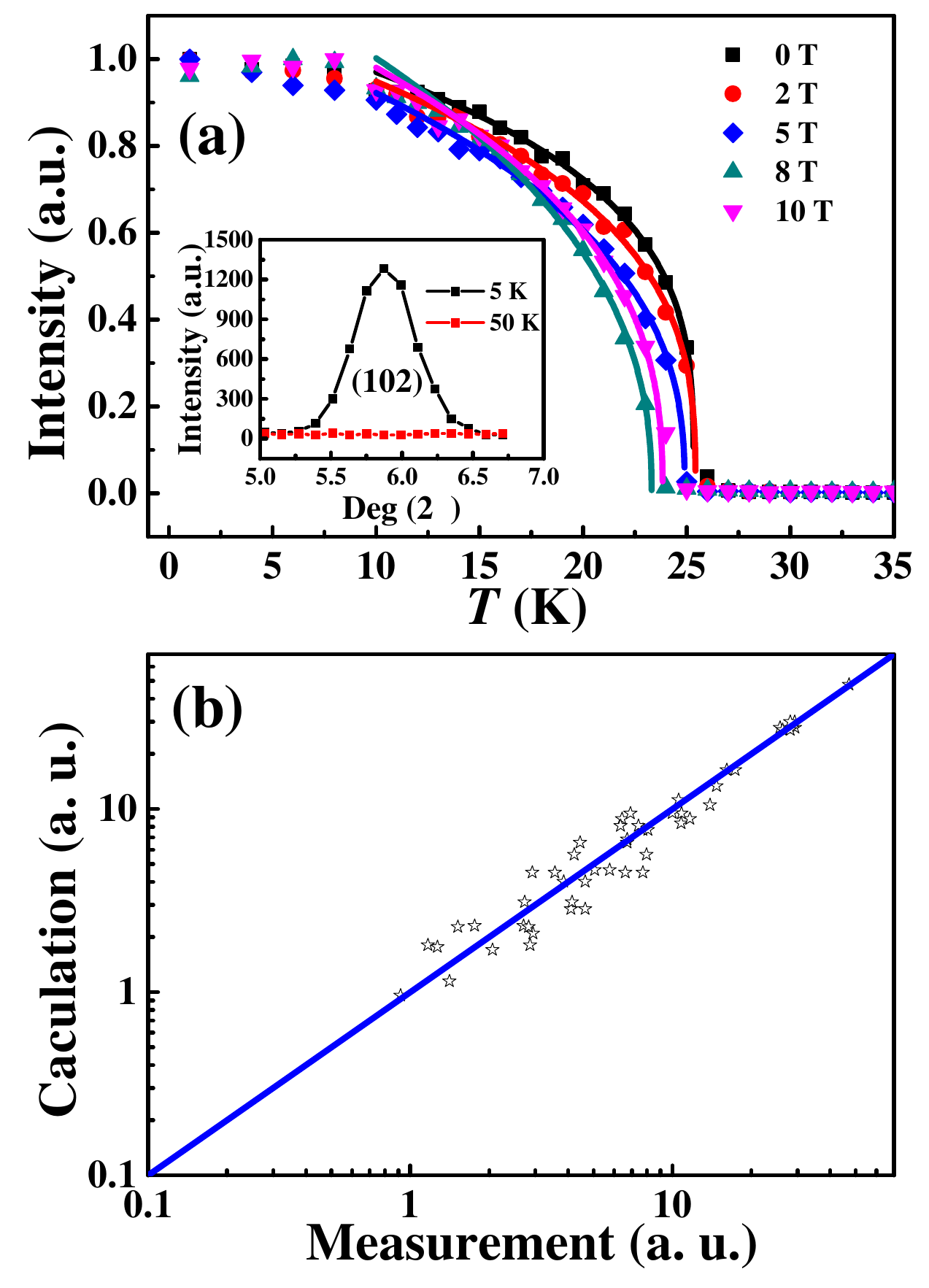}
	\end{center}
	\par
	\caption{(color online) (a) The temperature dependence of the normalized intensity for the (1 0 2) peak under different magnetic fields. The lines are fittings using the power law function described in the text. Insert:  the intensity of the (1 0 2) peak at 5 K and 50 K. The field is applied along the $b$-axis for BaMn$_2$Si$_2$O$_7$. (b) The comparison between the observed and calculated intensity for the magnetic Bragg peaks.
}
\end{figure}

To further probe the magnetic properties of BaMn$_2$Si$_2$O$_7$, the single crystal neutron diffraction  was performed under different temperatures and fields. As shown in the insert of Fig. 5(a), the intensity of the (102) peak appears at 5 K but disappears at 50 K. Since this peak is forbidden by the lattice symmetry, its appearance at 5 K represents the formation of an AFM ordering. The normalized intensity for the (102) peak, or the order parameter, under different fields are shown in Fig. 5. The critical temperature of the intensity, $T_{\text{N}}$, shifts to lower temperatures with increasing field H $\leq$ 8 T, and then shifts back to higher temperatures for 10 T data, which is consistent with the spin flop transition observed from other data listed above. The order parameter dates can be described by the power law function as

\begin{equation}
I=I_{0}\left(1-T/T_{N}\right)^{2\beta}
\end{equation}

 where $T_{\text{N}}$ is the ordering temperature, $I_0$ is the intensity at base temperature, and  $\beta$ is the order parameter critical exponent. The fitting of the data at different fields (solid lines in Fig. 5(a)) yields a value of $\beta$ ranging from 0.1413 to 0.1810 for different field, which is slightly smaller than the reported $\beta$ =  0.22 from polycrystalline sample studies. This small $\beta$ confirms the low dimensional nature of BaMn$_2$Si$_2$O$_7$.

\begin{figure*}[tbp]
\setlength{\abovecaptionskip}{-0.1cm}
	\linespread{1}
	\par
	\begin{center}
		\includegraphics[width= 7 in]{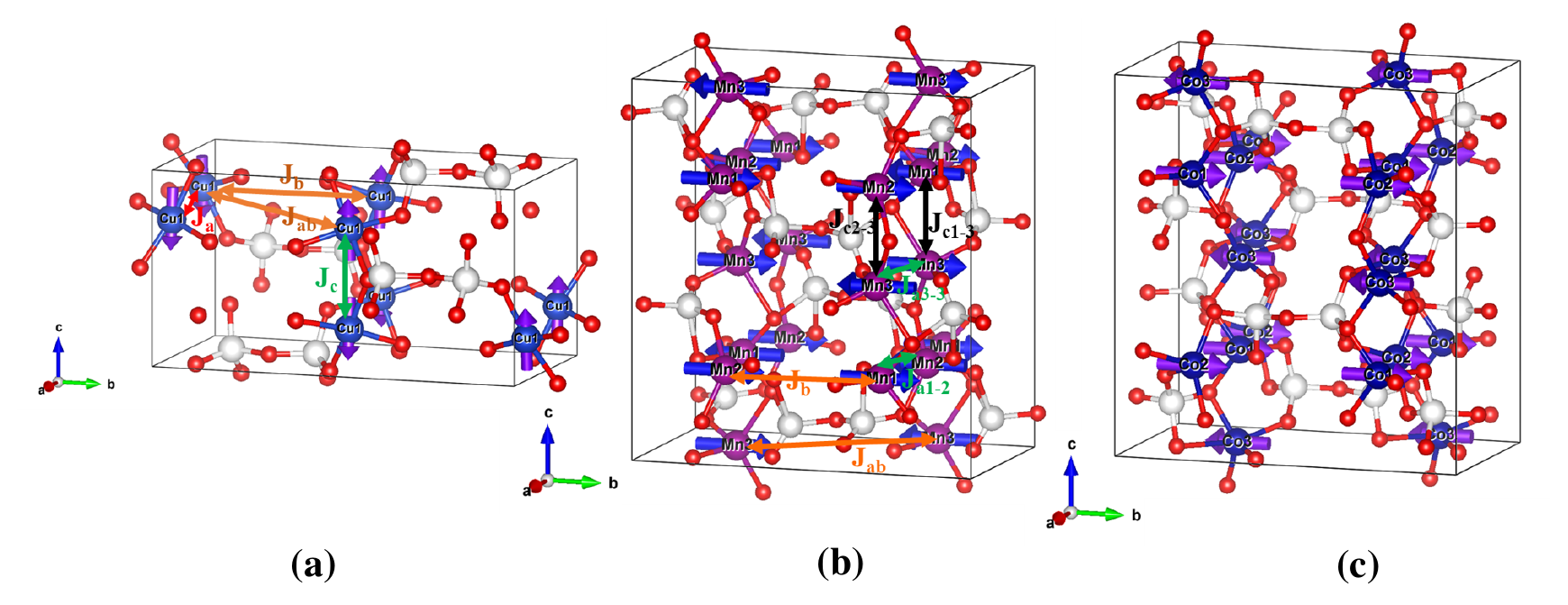}
	\end{center}
	\par
	\caption{(color online) The schematic superexchange paths and spin structures for BaM$_2$Si$_2$O$_7$. The white, red, and blue/pink spheres represent the Si$^{4+}$, O$^{2-}$, and M$^{2+}$ ions. The arrows represent the spins of M$^{2+}$ ions.
}
\end{figure*}

Totally, fifty-three magnetic Bragg peaks were observed. The refinement based on these peaks' intensity (Fig. 5(b)) yields a spin structure as plotted in Fig. 6(b). In this spin structure, the magnetic moment is along the $b$-axis, which is around 3.9 $\mu_{\text{B}}$ at 5 K. The arrangement of the spin along the $c$ and $a$-axis is antiferromagnetic, and along $b$-axis is ferromagnetic. This sin structure is slightly different from the reported one based on polycrystalline sample studies, in which the arrangement of the spin along the $a$-axis is ferromagnetic\cite{J. Ma}. Due to the average effect of powder samples, sometimes it will be difficult to exactly solve the spin structure based on polycrytalline data. Therefore, we prefer the spin structure obtained on this single crystal neutron diffraction study. Moreover, the refinement of the nuclear Bragg peaks yields the details of the crytallographic information, which is listed in Table I. There are some small discrepancies between this information  and the reported polycrystalline data\cite{J. Ma}. Again, we tend to believe the single crystal neutron data is more accurate.

\begin{table}
\par
\caption{Crystallographic information, including selected interatomic distances ({\AA}) and bond angle ({\degree}) of BaMn$_2$Si$_2$O$_7$ at 5 K. }
\begin{tabular}{p{2cm}<{\centering}p{1.5cm}<{\centering}p{3cm}<{\centering}p{1.5cm}<{\centering}}
\hline
\hline
\multicolumn{2}{l}{crystal symmetry}      &   monoclinic          \\
\multicolumn{2}{l}{space group}            &    C2/c           \\
\multicolumn{2}{l}{a (\AA)}  &                 7.2937           \\
\multicolumn{2}{l}{b (\AA)} &                 12.9656         \\
\multicolumn{2}{l}{c (\AA)} &                 13.9819         \\
\multicolumn{2}{l}{$\beta$ ({\degree})} &                 90.22     \\
\multicolumn{2}{l}{V (\AA$^3$)} &           1327.18   \\
 \hline
\multicolumn{4}{l}{along $c$ axis (along M chain)}                                   \\
 Mn(1)-O(4)         & 1.748       & Mn(1)-O(4)-Mn(3) & $105.22$       \\
Mn(2)-O(1)         & 2.058       & O(4)-Mn(3)-O(1)  & 120.83       \\
Mn(3)-O(1)         & 2.109       &  Mn(3)-O(1)-Mn(2) & 121.9        \\
Mn(3)-O(4)         & 2.477       &  O(1)-Mn(2)-O(1)  & 141.75       \\
                   &             &  O(4)-Mn(1)-O(4)  & 97.86        \\
\multicolumn{4}{l}{along $b$ axis {[}-Si(1)-O(5)-Si(2)-{]}}                        \\
Si(1)-O(5)         & 1.74        &  Si(1)-O(5)-Si(2) & 108.82       \\
Si(2)-O(5)         & 1.728       &                   &                    \\
\multicolumn{4}{l}{along $a$ axis  (between M chains)}                                                 \\
Mn(1)-O(2)         & 3.142       &                   &                    \\
Mn(2)-O(2)         & 1.974       &  Mn(2)-O(2)-Mn(1) & 89.1         \\
Mn(1)-O(6)         & 2.035       &  Mn(2)-O(6)-Mn(1) & 103.16       \\
Mn(2)-O(6)         & 2.642       &  Mn(3)-O(3)-Mn(3) & 99.37        \\
Mn(3)-O(3)         & 2.064       &                   &                    \\
Mn(3)-O(3)         & 2.565       &                   &                    \\
\multicolumn{4}{l}{distorted SiO$_4$ tetrahedra}                                                        \\
Si(1)-O(1)         & 1.565       &  O(1)-Si(1)-O(5)  & 105.26       \\
Si(1)-O(6)         & 2.08        &  O(1)-Si(1)-O(7)  & 136.29       \\
Si(1)-O(7)         & 1.485       & O(1)-Si(1)-O(6)  & 86.14        \\
Si(2)-O(2)         & 2.055       &  O(5)-Si(1)-O(6)  & 106.77       \\
Si(2)-O(3)         & 1.329       &  O(5)-Si(1)-O(7)  & 111.37       \\
Si(2)-O(4)         & 2.004       &  O(6)-Si(1)-O(7)  & 104.73       \\
                   &             &  O(4)-Si(2)-O(5)  & 94.59        \\
                   &             &  O(2)-Si(2)-O(4)  & 78.6         \\
                   &             & O(3)-Si(2)-O(4)  & 110.19       \\
                   &             & O(2)-Si(2)-O(3)  & 124.63       \\
                   &             &  O(2)-Si(2)-O(5)  & 108.72       \\
                   &             &  O(3)-Si(2)-O(5)  & 124.02       \\

\hline
\hline
\end{tabular}
\end{table}

\subsection{BaCo$_2$Si$_2$O$_7$}

\begin{figure}[tbp]
\setlength{\abovecaptionskip}{-0.1cm}
	\linespread{1}
	\par
	\begin{center}
		\includegraphics[width= 3.2 in]{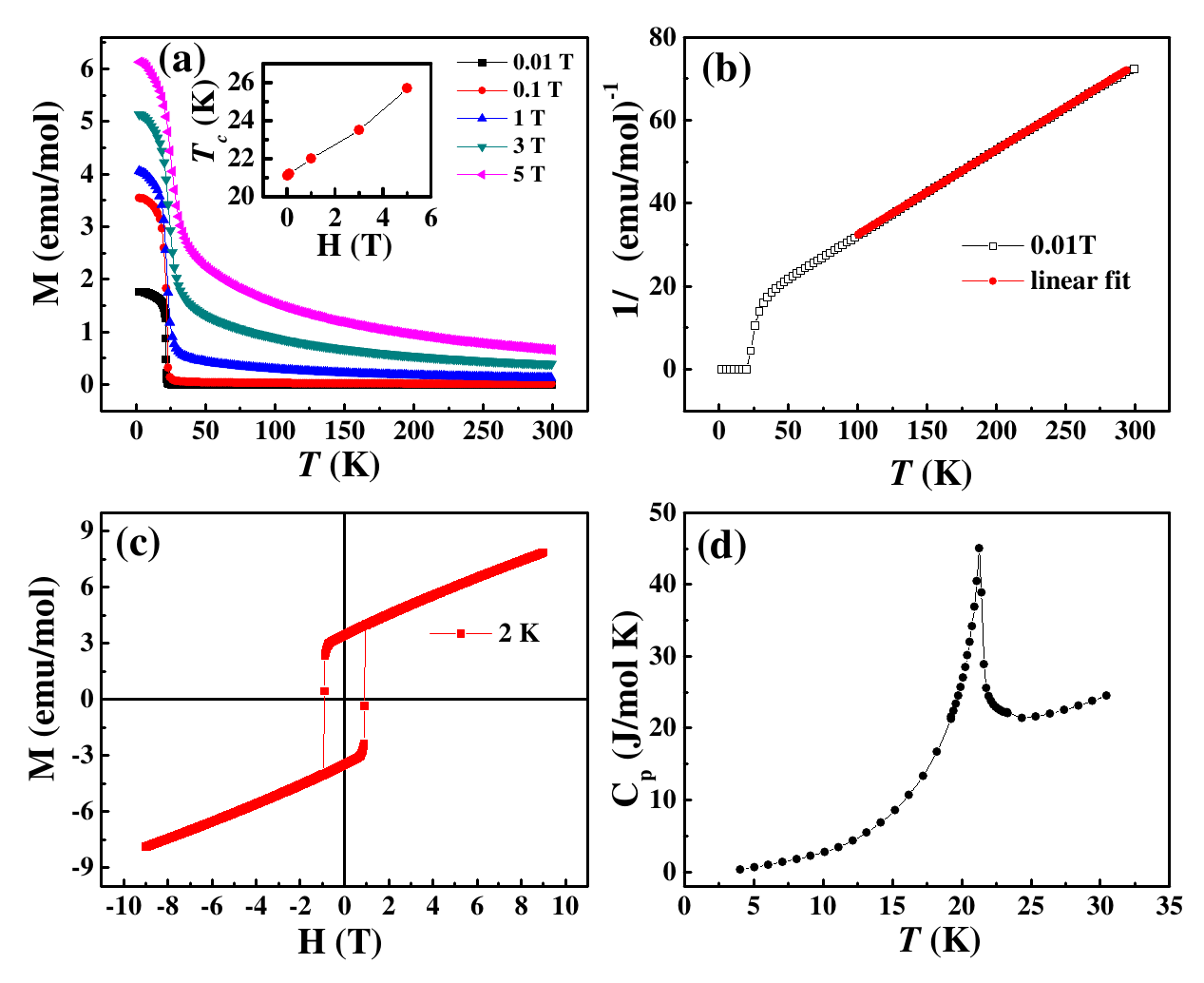}
	\end{center}
	\par
	\caption{(color online) All data for BaCo$_2$Si$_2$O$_7$. (a) The temperature dependence of magnetization under different fields. (b) The inverse of the susceptibility measured with H = 0.01 T. The squares are experimental data and the solid line is the linear fitting. (c) The hysteresis loop measured at 2 K. (d) The temperature dependence of the specific heat. Insert of (a): the field dependence of $T_{\text{C}}$.
}
\end{figure}

The magnetic properties of polycrystalline sample BaCo$_2$Si$_2$O$_7$ were also investigated. As shown in Fig. 7(a), the temperature dependence of magnetization exhibits a sharp increase below 21 K with H = 0.01 T. Meanwhile, the specific heat data (Fig. 7(d)) shows a sharp peak at 21 K. Both results consistently suggest a long range magnetic ordering occurring at 21 K. The transition temperature, $T_{\text{C}}$, is defined as the peak position of the derivative of magnetization (not shown here). As shown in the inset of Fig. 7(a), $T_{\text{C}}$ increases with increasing field. The hysteresis loop measured at 2 K (Fig. 7(c)) exhibits a square loop reaching a magnetic moment around 0.3 $\mu_{\text{B}}$/Co, which is still much smaller than the 3 $\mu_{\text{B}}$/Co, the full moment of spin-3/2 Co$^{2+}$ ions. All these behaviors, the increase of magnetization below $T_{\text{C}}$, the field-enhanced $T_{\text{C}}$, and the open loop all suggest a ferrimagnetic ground state for BaCo$_2$Si$_2$O$_7$. On the other hand, the Curie Weiss fitting from 100 to 300 K for the inverse of the susceptibility, as shown in Fig. 7(b), yields a Curie constant $\theta_{\text{CW}}$ = -58.5 K, which suggests AFM exchange interactions. Therefore, it is interesting for BaCo$_2$Si$_2$O$_7$ to have a ferrimagnetic ordering but with dominant AFM interactions.

\subsection{LSDA+U calculation}
To better understand the exchange interactions in the BaM$_2$Si$_2$O$_7$ (M = Cu, Co, and Mn) system, a LSDA+U calculation has been applied. In this calculation, the Heisenberg model was used by neglecting spin-orbital coupling and assuming the collinear moments. For Cu-sample, four interactions, $J_a$, $J_c$, $J_b$, and $J_{\text{[110]}}$ or $J_{ab}$ (Fig. 6(a)), has been taken into account. For Mn and Co-samples, there are three nonequivalent Mn/Co sites due to the lower lattice symmetry. Therefore,  six interactions $J_c$(M1-M3), $J_c$(M2-M3), $J_a$(M1-M2), $J_a$(M3-M3), $J_b$ , $J_{ab}$ (Fig. 6(b)) have been considered. Equations (2) and (3) are the Hamiltonians of M = Cu and Co/Mn samples, respectively. In equation (3), the n = 9 or 25 is the number of possible kinds of superexchanges for Co and Mn-sample, respectively.

\newcommand{\myvec}[1]%
{\stackrel{\raisebox{-2pt}[0pt][0pt]{\small$\rightharpoonup$}}{#1}}

\begin{equation}\label{e:barwq}
\begin{split}
H_{Cu}=&-(\frac{1}{4}{J_{c}\sum_{ij\in{c}}\myvec{s_{i}}\cdot{}\myvec{s_{j}}}+\frac{1}{4}{J_{b}\sum_{ij\in{b}}\myvec{s_{i}}\cdot{}\myvec{s_{j}}}\\&+\frac{1}{4}{J_{a}\sum_{ij\in{a}}\myvec{s_{i}}\cdot{}\myvec{s_{j}}}+\frac{1}{4}{J_{ab}\sum_{ij\in{ab}}\myvec {s_{i}}\cdot{}\myvec{s_{j}}})+H_{0}
\end{split}
\end{equation}

\begin{equation}
\begin{split}
H_{Co/Mn}=&-n(\frac{1}{4}{J_{c_{1-3}}\sum_{\mathclap{ij\in{c_{1}-c_{3}}}}\myvec{s_{i}}\cdot{}\myvec{s_{j}}}+\frac{1}{4}{J_{c_{2-3}}\sum_{\mathclap{ij\in{c_{2}-c_{3}}}}\myvec{s_{i}}\cdot{}\myvec{s_{j}}}\\&+\frac{1}{4}{J_{b}\sum_{ij\in{b}}\myvec{s_{i}}\cdot{}\myvec{s_{j}}}+\frac{1}{4}{J_{a_{1-2}}\sum_{\mathclap{ij\in{a_{1}-a_{2}}}}\myvec{s_{i}}\cdot{}\myvec{s_{j}}}\\&+\frac{1}{4}{J_{a_{3-3}}\sum_{\mathclap{ij\in{a_{3}-a_{3}}}}\myvec{s_{i}}\cdot{}\myvec{s_{j}}}+\frac{1}{4}{J_{ab}\sum_{ij\in{ab}}\myvec{s_{i}}\cdot{}\myvec{s_{j}}})+H_{0}
\end{split}
\end{equation}

The calculation was performed by using the Vienna $ab$ initio simulation package (VASP)\cite{Furthmller}.The experimental crystal structure was adopted in the calculation. For Cu and Co samples, the crystal structure data was adopted from Ref.\cite{Layland, Yamada}. For Mn sample, the structural data reported in this study was used. Several AFM spin structures were tested. By carefully varying U value, we successfully obtained lowest energy spin structures as same as the experimentally reported spin structures for the Cu and Mn samples, as plotted in Fig. 6(a) and Fig. 6(b). For Co-sample, we also listed our calculated ground state as shown in Fig. 6(c). The used parameter for calculations are U = 5.6 eV\cite{Bertaina},  J = 1 eV for BaCu$_2$Si$_2$O$_7$, U = 7 eV, J = 1 eV for BaMn$_2$Si$_2$O$_7$, and U = 5 eV, J = 1 eV for  BaCo$_2$Si$_2$O$_7$. Their magnetic moment per magnetic atom in cell is 0.6, 4.6 and 2.6 $\mu_{\text{B}}$ for Cu, Mn, and Co samples, respectively, which were obtained from the results of ions relaxation. The calculated exchange interactions are listed in Table II.

\begin{table*}[]
\par
{\caption{ $J$ values(meV) deduced from the LSDA+U calculations. Negative J represents AFM interaction and positive J represents FM one. }}
\label{t1}
\setlength{\tabcolsep}{2.6mm}{
\begin{tabular}{ccccccc}
\hline
\hline
       & \multicolumn{2}{c}{BaCu$_2$Si$_2$O$_7$}                                                                             & \multicolumn{2}{c}{BaCo$_2$Si$_2$O$_7$}         & \multicolumn{2}{c}{BaMn$_2$Si$_2$O$_7$}         \\ \hline
 & LSDA+U & \begin{tabular}[c]{@{}c@{}}Mean-field$^{11}$  and Spin wave$^{17}$\end{tabular} & \multicolumn{2}{c}{LSDA+U}        & \multicolumn{2}{c}{LSDA+U}        \\ \hline
$J_c$     & -17.6    & -24.1                                                                 & -0.40 (Co3-Co3) & -0.40 (Co1-Co2) & -0.13 (Mn3-Mn3) & -0.13 (Mn1-Mn2) \\
$J_b$     & 0.18     & -0.2                                                                  & \multicolumn{2}{c}{0.03 }           & \multicolumn{2}{c}{2E-4}               \\
$J_a$     & 0.27     & 0.46                                                                  & 0.10 (Co3-Co3)  & 0.10 (Co1-Co2)  & -0.03 (Mn3-Mn3) & -0.04 (Mn1-Mn2) \\
$J_{ab}$    & -0.08    & -0.15                                                                 & \multicolumn{2}{c}{0.01 }           & \multicolumn{2}{c}{-2e-4}              \\ \hline
\hline
\end{tabular}}
\end{table*}

Several noteworthy features for the exchange interactions are: (i) for Cu sample, the dominating interaction is $J_c$ = -17.6 meV, which is at least two orders of magnitude larger than the other interactions. Most of the obtained interactions here are pretty comparable to those obtained from the mean field theory calculation \cite{Tsukada}and experimental spin wave measurement\cite{Kenzelmann}, as listed in Table II. Only for $J_b$, we obtained 0.18 meV and the spin wave simulation leads to -0.2 meV. This comparison further validates our calculations; (ii) the $J_c$ of Cu sample is much larger than those for Mn ($J_c$ = -0.13 meV) and Co ($J_c$ = -0.40 meV) samples; (iii) for Mn and Co samples, the $J_c$ is not significantly larger than $J_a$ and they are comparable on energy scale; (iv) the $J_a$ = 0.1 meV for Co sample is in FM nature but it is in AFM nature  for Mn  sample as -0.03 $\sim$ -0.04 meV; (v) for all samples, the $J_b$ and $J_{ab}$ are the weakest interactions.

\section{DISCUSSION}

\begin{figure}[tbp]
\setlength{\abovecaptionskip}{-1mm}
	\linespread{1}
	\par
	\begin{center}
		\includegraphics[width= 3.2 in]{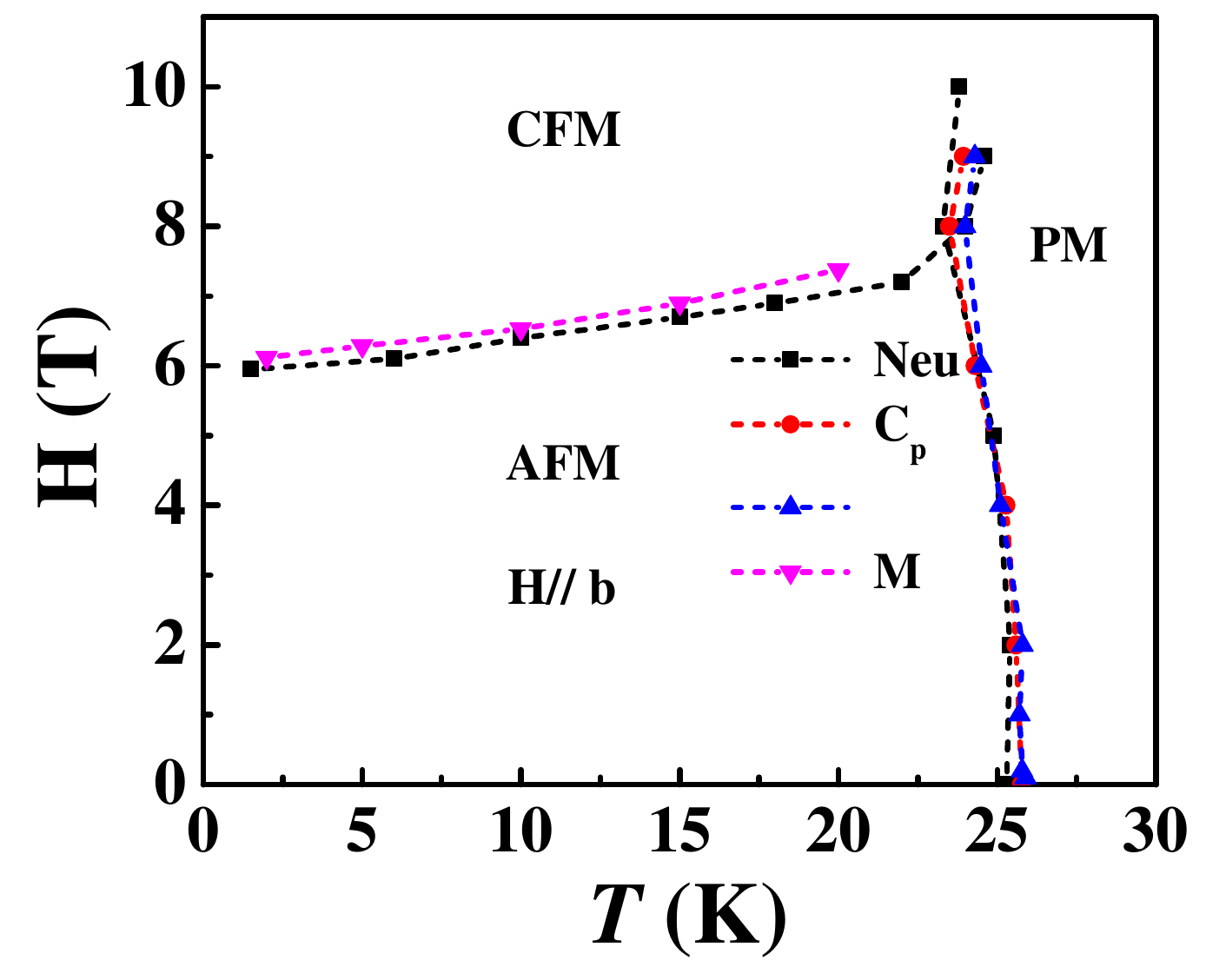}
	\end{center}
	\par
	\caption{(color online) Magnetic phase diagram for BaMn$_2$Si$_2$O$_7$.
}
\end{figure}

First, we present the magnetic phase diagram for BaMn$_2$Si$_2$O$_7$ with H along the $b$-axis, which was constructed by combining the susceptibility, magnetization, heat capacity and neutron diffraction data, as shown in Fig. 8. At zero field, a long range antiferromagnetic ordering is  formed below $T_{\text{N}}$ = 26 K. With applied field,  $T_{\text{N}}$ shifts to lower temperatures with increasing field. Around 6 T, there is a spin flop transition. Since the applied field along the $b$-axis should induce the spins to flop to its perpendicular direction, which could be $ac$ plane and also since between 6 $\sim$ 9 T, the magnetization keeps linearly increasing, the spin state after the spin flop could be a canted antiferromagnetic state in $ac$ plane. Its exact nature needs future studies to clarify. It is noteworthy that a spin flop transition also occurs in BaCu$_2$Si$_2$O$_7$ with an applied field along the $c$-axis and with a smaller critical field around 2 T. As shown in Fig. 6, the spin structure is different between the Cu and Mn samples, and therefore it is not surprised to see this different  easy axis for spin flop transition.

Second, we want to discuss the exchange interactions of BaM$_2$Si$_2$O$_7$ in more details. For Cu sample, the interaction between the nearest Cu$^{2+}$ is through superexchange Cu-O-Cu along the $c$ axis but super-superexchange through  Cu-O-Si-O-Cu along the $a$ axis and Cu-O-Si-O-Si-O-Cu along the $b$ axis. Therefore, its reasonable to see our calculation lead to dominating $J_c$, which further confirming that Cu-sample is approaching a one dimensional nature for magnetism.

The fast drop of $J_c$ from Cu to Mn/Co samples could be related to their different outshell electronic configurations. For Cu$^{2+}$ ion with one orbital half-filled, the interaction along $c$ axis only depends on the superexchange of 3d$_{x^2-y^2}$ (Cu$^{2+}$)-p(O$^{2-}$)-3d$_{x^2-y^2}$ (Cu$^{2+}$). For Mn$^{2+}$ ion,  its five half-filled 3d orbitals split to three high energy $t_{\text{2g}}$ and two low energy $e_{\text{g}}$ orbitals due to the MnO$_4$ tetrahedral site. All five orbitals can involve super-exchange interaction through p(O$^{2-}$) orbitals with orbitals in another Mn$^{2+}$ ion, which results to 25 kinds of combination of superexchange in Mn compound. Since the bond angles, 121.9 $^\circ$ for Mn3-O1-Mn2 and 105.22 $^\circ$  for Mn1-O4-Mn3 along the $c$-axis are between 90$^\circ$ and 180$^\circ$, each interaction of these 25 combination is determined by the competition between ferromagnetic and anti-ferromagnetic superexchange nature according to GKA rules\cite{Goodenough}. Then it is possible that this competition can compensate each other within the possible interactions and lead to a significant decrease of the total $J_c$ interaction.  For Co-sample with isostructure to Mn sample, there are three half-filled 3d orbitals, similar situation can occur too and therefore again leads to smaller magnitude of $J_c$.

The structural distortion in  Mn/Co sample has another consequence, which is that the ratio between the  M-M bond length along the $c$ and $a$-axis for Mn/Co now is smaller than that for Cu-sample. Plus the fact that  $J_c$ is reduced in Mn/Co sample, the $J_c$ and $J_a$ now are comparable and therefore make the Mn/Co sample approaching three dimensional nature in magnetism. This could be the reason for the higher magnetic ordering temperatures in Mn/Co samples, comparing to that of Cu sample. It is also interesting to observe different magnetic ground states for Mn and Co samples with isostructure, which is AFM and ferrimagnetic-like, respectively. This difference reflected in our calculation is the different nature of $J_a$, which is in AFM nature for Mn but FM nature for Co sample. For Co$^{2+}$ ion there is no half filled $e_{\text{g}}$ orbitals attending the interactions but Mn$^{2+}$ ion has two half filled $e_{\text{g}}$ orbitals involved. It is possible that the interactions involving these $e_{\text{g}}$ orbitals along the $a$-axis favor AFM nature and meanwhile the total interactions of Co sample are still under competition between AFM and FM nature as argued above. Therefore, the reduction of the AFM interaction part due to the lack of half filled $e_{\text{g}}$ orbitals can lead to a final FM nature of the total interaction for Co sample. Our calculated spin structure for Co-sample indeed shows FM arrangement of the spin along the $a$ and $b$-axis AFM arragnement along the $c$-axis. Since our calculation assumed the collinear spins, this calculated spin structure of the Co-sample could be slightly different from the experimental result. While its true spin structure needs further neutron measurements to be confirmed, the magnetization measurements on single crystal samples\cite{Layland} suggested that the spins are canted along the $c$-axis with a small angle around 5 $^\circ$ and this canting should be the origin for its ferrimagnetic-like magnetization.

\begin{figure}[tbp]
\setlength{\abovecaptionskip}{-0.1cm}
	\linespread{1}
	\begin{center}
		\includegraphics[width= 3.2 in]{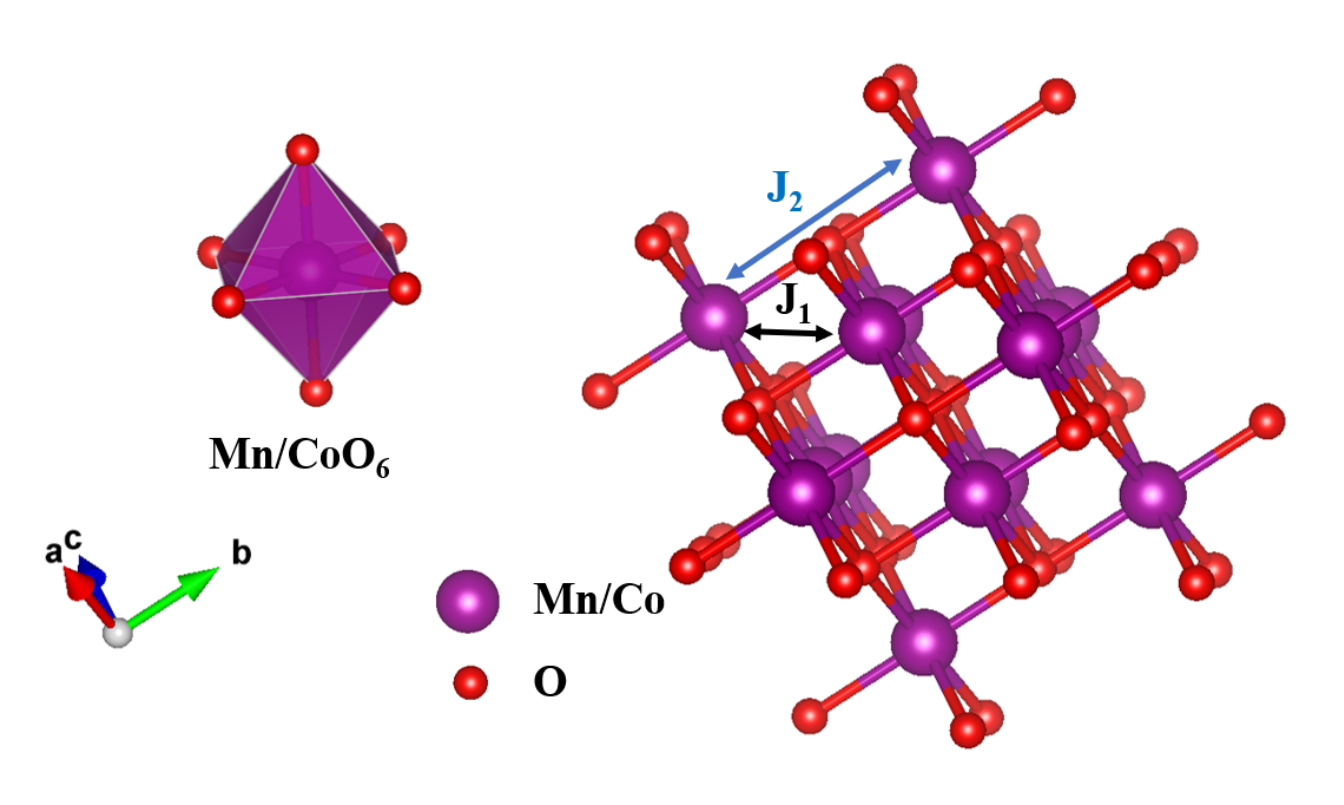}
	\end{center}
	\caption{(color online) Schematic crystal structure for Mn/CoO. The red and pink spheres represent the O$^{2-}$ and Mn/Co$^{2+}$ ions, respectively.
}
\end{figure}
\begin{table}[]
\caption{Interaction values(meV) of MnO, CoO in previous work.}
\resizebox{0.48\textwidth}{!}{
\begin{tabular}{cccccc}
\hline
\hline
   & \begin{tabular}[c]{@{}c@{}}MnO(cal.)$^{33,34}$\\Energy-bond theory\end{tabular}    & \begin{tabular}[c]{@{}c@{}}MnO(cal.)$^{35}$\\AB initio\end{tabular}        & \begin{tabular}[c]{@{}c@{}}MnO\\(Exp.)$^{36}$ \end{tabular} & \begin{tabular}[c]{@{}c@{}}CoO(cal.)$^{35}$\\AB initio\end{tabular}       & \begin{tabular}[c]{@{}c@{}}CoO\\(Exp.)$^{37}$\end{tabular}     \\
   \hline
$J_1$ & \multirow{2}{*}{-0.98 } & \multirow{2}{*}{-0.91 } & -2.06 ,                                                      & \multirow{2}{*}{-0.32 } & \multirow{2}{*}{0.7 } \\
   &                           &                           & -2.64                                                        &                          &                         \\
$J_2$ & -1.74                   & -1.99                   & â-2.79                                                        & -4.84                  & -6.3     \\
\hline
\hline
\end{tabular}}
\end{table}

To further elaborate this interesting difference between Mn and Co samples, we look into another case, the monoxide MnO and CoO. This system is with a NaCl structure with M$^{2+}$ ion on the MO$_6$ octahedral site. As shown in Fig. 9, the major exchange interactions are the $J_1$ between the nearest neighbor M$^{2+}$ ions with 90 $^\circ$ M-O-M bond angle and the $J_2$ between the next nearest neighbor M$^{2+}$ ions with 180 $^\circ$ M-O-M bond angle. As listed in Table III, the reported experimental and theoretical values \cite{Oguchi,Oguchi2,Fischer,Pepy,Tomiyasu} of $J_2$ are in AFM nature for both MnO and CoO. For $J_1$, the experimental value is in AFM nature for MnO but FM nature for CoO. Meanwhile, the theoretical values of $J_1$ are in AFM nature for both MnO and CoO but with a smaller magnitude for CoO. Despite the inconsistency between the experimental and theoretical values of $J_1$ for CoO, one general trend is that the AFM nature of $J_1$ for CoO is suppressed by comparing to that of MnO. This trend is similar to the $J_a$ case for BaMn$_2$Si$_2$O$_7$ and BaCo$_2$Si$_2$O$_7$.  For a octahedral site, the 3d orbitals split to three low energy $t_{\text{2g}}$ and two high energy $e_{\text{g}}$ orbitals. In MnO/CoO, the superexchange direction involving $t_{\text{2g}}$ orbitals along $J_1$ is similar to those of $J_a$ in BaM$_2$Si$_2$O$_7$. It is possible that the lack of half filled $t_{\text{2g}}$ orbitals for CoO leads to the reduction of AFM interaction and therefore leads to a total FM nature of $J_1$ if we assume the experimental value is more trustable, which is similar to the reduction of AFM interaction due to lack of half filled $e_{\text{g}}$ orbitals in BaCo$_2$Si$_2$O$_7$.

 \section{CONCLUSION}
In summary, we constructed a magnetic phase diagram and obtained more accurate antiferromagnetic spin structure for single crystalline BaMn$_2$Si$_2$O$_7$ by performing susceptibility, magnetization, specific heat and single crystal neutron diffraction measurements. We also confirmed the ferrimagnetic-like property for polycrystalline BaCo$_2$Si$_2$O$_7$. Finally, in order to systematically understand the  different magnetic ground states in the  BaM$_2$Si$_2$O$_7$ (M = Cu, Co, and Mn) system, we performed LSDA + U calculations and extracted their magnetic exchange interactions. Our discussions based on the comparison among the exchange interactions suggest that the structure distortion and different outshell electronic configuration are the main reasons for the changes of magnetic properties from Cu to Co and to Mn samples.

\begin{acknowledgments}
This work is supported by the Natural Science Foundation of China(NSFC) with grant number 11774223. J. M., G. H. W. and H. D. Z. acknowledge the support from the Ministry of Science and Technology of China(2016YFA0300500). Q. H. thanks the support from the GO Student Program of ORNL.  H. D. Z. also thanks for the support from NSF-DMR with grant number NSF-DMR-1350002.
\end{acknowledgments}

\end{document}